\documentclass[12pt]{spieman}  
\usepackage{amsmath}
\usepackage{amsfonts}
\usepackage{amssymb}
\usepackage{graphicx}
\usepackage{setspace}
\usepackage{tocloft}
\usepackage{hhline}
\usepackage{multicol}
\usepackage[left=1.65cm, right=1.65cm,top=1.5cm, bottom=1.5cm]{geometry}

	\usepackage{fancyhdr}

\usepackage{wasysym}

\newcommand{\nuvu}{N\"uv\"u } 

\usepackage{amsmath}

\title{End-to-end ground calibration and in-flight performance of the FIREBall-2 instrument}

\author[a, b]{Vincent Picouet}
\author[a]{Bruno Milliard}
\author[c,d]{Gillian Kyne}
\author[a]{Didier Vibert}
\author[b]{David Schiminovich}
\author[c]{Christopher Martin}
\author[e]{Erika Hamden}
\author[c]{Keri Hoadley}
\author[f]{Johan Montel}
\author[b]{Nicole Melso}
\author[c]{Donal O'Sullivan}
\author[f]{Jean Evrard}
\author[f]{Etienne Perot}
\author[a]{Robert Grange}
\author[d]{Shouleh Nikzad}
\author[a]{Philippe Balard}
\author[a]{Patrick Blanchard}
\author[f]{Frederi Mirc}
\author[f]{Nicolas Bray}
\author[d]{April Jewell}
\author[a]{Samuel Quiret}
\affil[a]{Aix Marseille Univ, CNRS, CNES, LAM, Marseille, France}
\affil[b]{Department of Astronomy, Columbia University, 550 W. $120^{th}$ Street, New York, NY 10027, USA}
\affil[c]{Cahill Center for Astrophysics, California Institute of Technology, 1216 East California Boulevard, Mail Code 278-17, Pasadena, CA 91125, USA}
\affil[d]{Jet Propulsion Laboratory, California Institute of Technology, 4800 Oak Grove Drive, Pasadena, CA 91109, USA }
\affil[e]{University of Arizona, Steward Observatory, 933 N Cherry Ave, Tucson, AZ 85721, USA}
\affil[f]{Centre National d'Etudes Spatiales, 31401 Toulouse CEDEX 4, France}

\cftpagenumbersoff{figure}
\cftpagenumbersoff{table} 
\begin{document} 
\maketitle

\begin{abstract}
The payload of the Faint Intergalactic Redshifted Emission Balloon (FIREBall-2), the second generation of the FIREBall instrument (PI: C. Martin, Caltech), has been calibrated and launched from the NASA Columbia Scientific Balloon Facility (CSBF) in Fort Sumner, NM. FIREBall-2 was launched for the first time on the $22^{\textrm{nd}}$ September 2018, and the payload performed the very first multi-object acquisition from space using a multi-object slit spectrograph (MOS).
This performance-oriented paper presents the calibration and last ground adjustments of FIREBall-2,  the in-flight performance assessed based on the flight data, and the predicted instrument's ultimate sensitivity. 
This analysis predicts that future flights of FIREBall-2 should be able to detect the HI Ly$\alpha$  resonance line in galaxies at $z \sim 0.67$, but will find it challenging to spatially resolve the circumgalactic medium (CGM). \end{abstract}


{\noindent \footnotesize\textbf{*}Vincent Picouet,  \linkable{vincent.picouet@lam.fr} }



\begin{spacing}{1}   

\begin{multicols}{2}

\section{INTRODUCTION} \label{sec:intro}  

The circumgalactic medium (CGM) plays a critical role in the evolution of galaxy discs, as it hosts important mechanisms regulating their replenishment through inflows and outflows.\cite{Tumlinson2017}
Directly mapping the HI Lyman alpha (Ly$\alpha$) emission from circumgalactic gas at high redshifts ($z>2$) is an insightful alternative to absorption spectroscopy, providing a new perspective with a complete 2-D or 3-D mapping of the gas distribution\cite{Martin2014,OSullivan2019,Wisotzki2015,Cantalupo2014,Battaia2018,Martin2019
}. Its importance notwithstanding, emission data are very scarce in the $10$-billion-year span from $z\sim2$ to the present because of the difficulties inherent to vacuum UV observations. 

The FIREBall-2 instrument, jointly funded by CNES and NASA, has been developed to help fill this gap and pave the way for future orbital projects. It has been optimized to provide a bi-dimensional (x, $\lambda$) map of the extremely faint diffuse Ly$\alpha$ HI emission in the CGM at $z\sim0.7$ and can to observe around 200 targets in a single night's flight.

FIREBall-2 is designed to have a wide field of view (FOV) of $20.5$x$37$ arcmin$^2$,  a sharp angular resolution of $\sim 6$ arcsec FWHM and a resolving power of $\sim 1500$ for a diffuse object. Given these instrument design choices, FIREBall-2 has the capability to make direct measurements and set statistical constraints on the mass, distribution and velocity of multi-phase gas in the CGM. FIREBall-2 targets multi-object high-resolution spectra in order to simultaneously address the following science goals: measuring the total energy budget of the CGM, studying the gas distribution in galaxy halos as a function of galaxy type, constraining inflow and outflow mechanisms, deriving budgets for the exchange of mass, metals, and energy between galaxies and their surroundings, as well as many other applications.

The instrument is a balloon-borne one-meter telescope coupled to a UV multi-object spectrograph (MOS) designed to image the CGM in emission. Its goal is to target the Ly$\alpha$ line redshifted in the stratospheric balloon flight's spectral window near $200$ nm for the universe at $z\sim0.7$.

The overall science requirements set ambitious technical goals within a budget of a stratospheric program in terms of detector, optics, and guidance. The instrument design choices along with their technical rationale are summarized in Table \ref{FIREBall's design summary}.

In this paper, we describe the end-to-end ground calibration and in-flight performance of the FIREBall-2 instrument.  In Section \ref{sec2} we describe the instrument design and summarize the last ground adjustements of FIREBall-2. The reduction pipeline of the flight data is described in Section \ref{sec3} and the in-flight performance assessed based on the available flight data is described in Section \ref{sec4}. As FIREBall-2 has been funded for two other flights, we will explore in Section 5 the extrapolated instrument's optimal sensitivity  based on flight and post-flight data.

\section{Instrument overview and calibration}\label{sec2}

\subsection{FIREBall-2 design summary}\label{sec21}
The FIREBall-2 gondola has a height of $\sim 5$ meters hand a weight of $\sim 2.3$ metric tons (including the $500$-kg ballast)\cite{hamden2020}. This payload is designed to reach an altitude of 35 km attached to a stratospheric balloon filled with $1.1$ million cubic meters of helium. Because the atmosphere absorbs most of the UV light coming from the cosmos,  the FIREBall spectrograph has been optimized to operate in the narrow atmospheric transmission window around $2000 $  \AA{} where there is a dip in the atmospheric UV absorption above $35$ km\cite{Matuszewski2012}. 
The nominal UV band-pass for this window ($1990$ - $ 2130 $ \AA{}) transmits nearly half of the incident UV light and the residual continuum atmospheric background is comparable to or lower than the extragalactic UV background. Within this atmospheric window, there is a contaminating emission band between $2044$ - $2062$ \AA{} from NO-$\delta$, which is emitted at altitudes well above that of the balloon, but the zodiacal contribution is negligible \cite{Leinert1998}.  The typical atmospheric transmission is $50\%$ when the balloon is at float altitude and the pointing has an elevation of $50^\circ$. In the subsequent modeling, we have assumed a total sky background (extragalactic $+$ atmospheric) of $500$ continuum units\footnote{Continuum Units : $n_{photon} \cdot s^{-1} \cdot cm^{-2} \cdot sr^{-1} \cdot \mathring{A}^{-1}$}.

\begin{figure*}
\centering
\includegraphics[width=\textwidth]{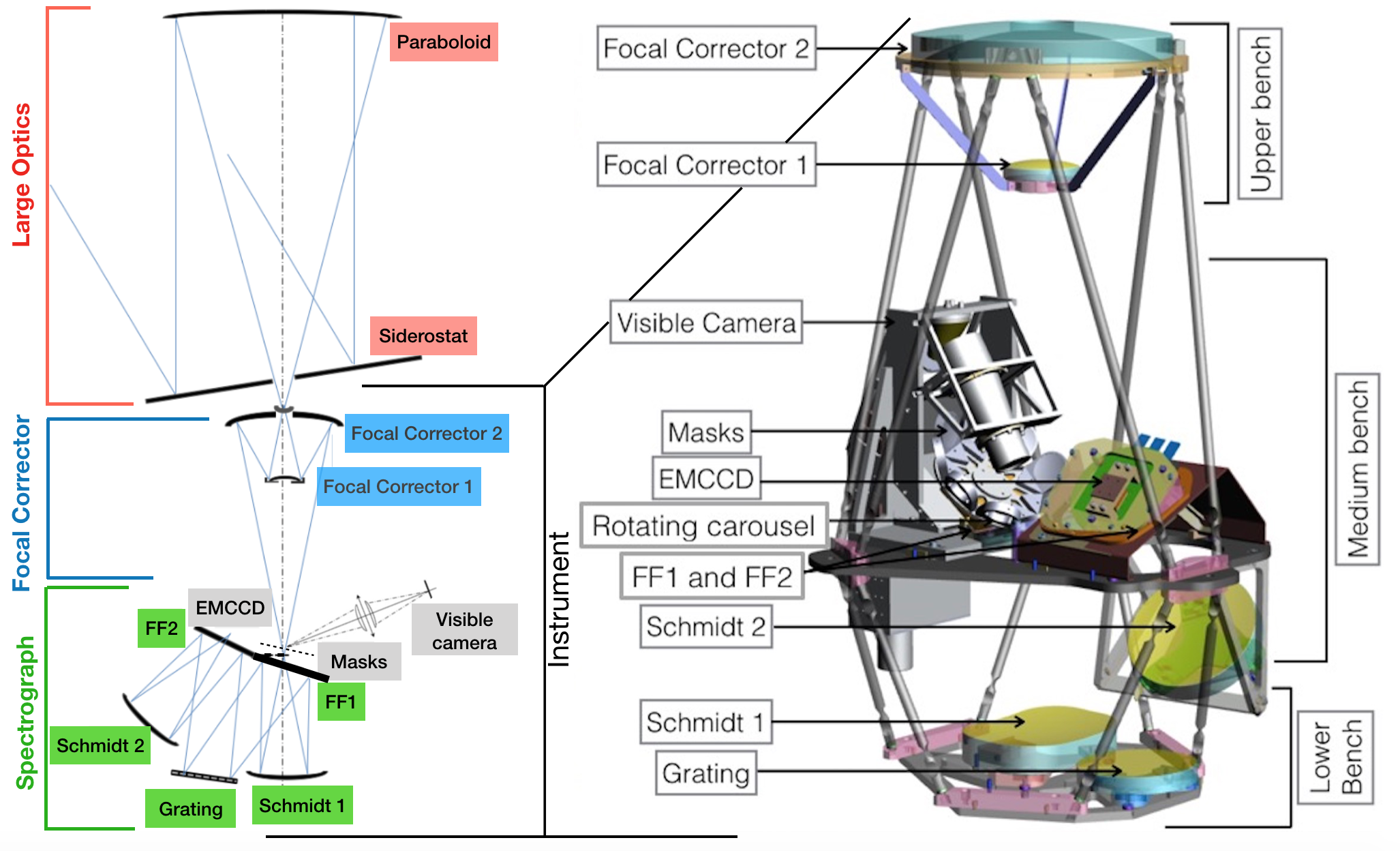}\\
\caption{\textbf{Left}: FIREBall-2 optical design layout. For readability, the instrument is shown in the payload layout as a flattened, plane optical layout, rather than the compact 3-D design in the real world.
\textbf{Right}: Interior of the FIREBall-2 spectrograph. Note that it is rotated of about 180 degrees along the z axis compared to the left side in order to improve the visualization of the different components. 
During flight this equipment is in a vacuum tank kept at $\sim 10^{-6}$ mB by cryopumping and cryosorbtion at $\sim100$K which reduces heat losses in the detector cooling chain by minimizing ice trapping, and cools the detector near $170$K. The whole vacuum tank is placed on tip-tilt actuators in order to control the focus and centering during the flight.}
\label{design}
\end{figure*}

The optical design of FIREBall-2 relies on a low-inertia $1.2$-meter sidereostat mounted on a controlled gimbal system that stabilizes the reflected beam in the gondola's frame and directs it to a fixed $f/2.5$ paraboloid that in turn focuses it at the entrance of the $f/2.5$ instrument (see Figure \ref{design}). 
The object selection is achieved with a series of pre-installed precision slit mask systems that also feed the guider camera (see Figure \ref{MGS}). The detector is a  \textit{Teledyne-e2v} electron multiplying CCD (EMCCD) anti-reflection-coated and delta-doped by the Jet Propulsion Laboratory (JPL) \cite{Nikzad} and Caltech.

The one-meter telescope shows very good performance at a cost compatible with suborbital programs achieved by using an innovative thin mirror/hierarchical-kinematic support and telescope pointing control.\cite{Montel}
The telescope is mounted on a low-mass gondola made of aluminum and carbon-fiber. This gondola provides four-axis guidance (a coarse azimuth control for the gondola to face the field, fine elevation and cross-elevation to target the exact center of the field despite gondola pendulum, and field rotation)  over a range of $40$ - $70$ degrees in elevation. During the 2018 flight, in degraded conditions (see Section \ref{sec31}), the pointing stability performance on each axis was $\sim0.5$ arcsec.

Inside the tank, The narrow field of view of the parabola alone is extended to $\sim800$ arcmin$^2$ by a Schwartzchild-type two-mirror field corrector $f/2.5-f/2.5$  (Figure \ref{design}). The corrector is designed to introduce a field curvature that cancels the native curvature of the Schmidt-based spectrograph to produce a flat focal surface at the detector. The slits masks of the spectograph have been fabricated at the same curvature and are mounted on a rotation stage. The spectrograph itself consists of identical $f/2.5$ Schmidt collimator and camera mirrors, the aspheric correction being provided by the $2400$ g/mm reflection grating obtained by double replication  of a deformable matrix using the Lemaitre method \cite{Lemaitre2014}. Flat folding mirrors are used to reduce the volume of the vacuum tank in which all the optics and detector are encased. 
The Invar spectrograph structure and the Zerodur (TM) low-expansion glass make the instrument robust against temperature changes. A layout of the overall payload and the instrument are respectively presented in Figure \ref{design}.

\begin{table}
\centering 
\caption{FIREBall's design summary}

\renewcommand{\arraystretch}{1.4}
\label{FIREBall's design summary}
\begin{tabular*}{\linewidth}{@{\extracolsep{\fill}}p{0.20\linewidth}p{0.32\linewidth}p{0.45\linewidth}@{}}
\hline
\textbf{Parameter} & \textbf{Value}  & \textbf{Rationale}  \\ 
\hhline{===}
\textbf{Telescope Design} & $1.25$-m moving flat siderostat. Fixed $f/2.5$  paraboloid ($1.0$ m) & Siderostat based pointing system reduces moving parts inertia and total cost - $f/2.5$ maximizes the FOV and diffuse sensitivity\\ 
\textbf{Focal corrector} &  f/2.5 two mirror system  &  provides a corrected $0.6^\circ$ FOV  \\ 

\textbf{Spectrograph} & Fast UV Multi Object Spectrograph with an aspherical grating &  More efficient than IFU for the low density bright enough targets  \\ 
\textbf{Detector} & Delta-doped EMCCD in photon-counting mode & Read noise $\sim0$ and increased QE  ($\sim50\%$) for photon-starved astronomy  \\ 
\textbf{Guiding system} & $33$Hz - $1.0$-meter pupil in visible & guide with   $\epsilon_{\textrm{Tracking}}<1''$ rms/axis on stars $<12^{\textrm{th}}$ mag\\ 
\textbf{Wavelength range} & $199 - 213$nm & stratospheric UV window  \\ 
\textbf{Cooling system} & Ice fusion and water boiling system & Dissipate 35MJ in the $[-60,+40]^\circ C$ range of environment T$^\circ$\\ 
\textbf{Mask system} & 4 science masks & Adapted to observation time\\
 & 5 calibration masks & Required for calibration procedures\\
  
\textbf{Spatial resolution} &   $\sim6''$ in both axis as the PSF is&  Resolve CGM emission at $z\sim0.7$ and\\
 (\textbf{FWHM})&hardly affected by the slit width & separate it from disk emission \\ 

\textbf{Resolving power } & R$_{\textrm{diffuse}}\sim1500 \lambda/d\lambda$,   & Resolve gas kinematics ($> 10^2$ km/s)\\
\textbf{ } & R$_{\textrm{Pointlike}}\sim2000 \lambda/d\lambda$ & 
\\ 
\textbf{Slit dimensions} &slit$_{\textrm{dim}} \sim 6'' \times 24''$ & Target CGM extension\\
\textbf{} & slit$_{\textrm{width}} \sim 1.5$ \AA{}    & Based on SNR optimization\\

\hline 

\end{tabular*} \\

\end{table}
In addition to the science objectives described in the introduction, the flight of the FIREBall-2 instrument also validated many processes and state-of-the-art hardware components (UV EMCCD \cite{kyne2020deltadoped}, CNES pointing system \cite{Montel2019}, cooling system, large UV aspherical grating \cite{Lemaitre2014}).

\subsection{Main ground alignment}

The FIREBall-2 instrument complexity and the intricacy of its sub-components require a hierarchic adjustment and calibration procedure. An especially demanding specification is the absolute guidance to 1 arcsec ($\sim 1/6$ slit width) during the $\sim 2$ hours of observation per field. This goal is made still more challenging by the slight mask-to-mask shape differences  (see section \ref{sec21}) that require a mask-dependent calibration.\\

The optical fine adjustment procedure has been split into two major blocks, each containing several alignment procedures. \\
\subsubsection{Internal Adjustments}

\begin{figure*}
\centering
\includegraphics[width=1\textwidth]{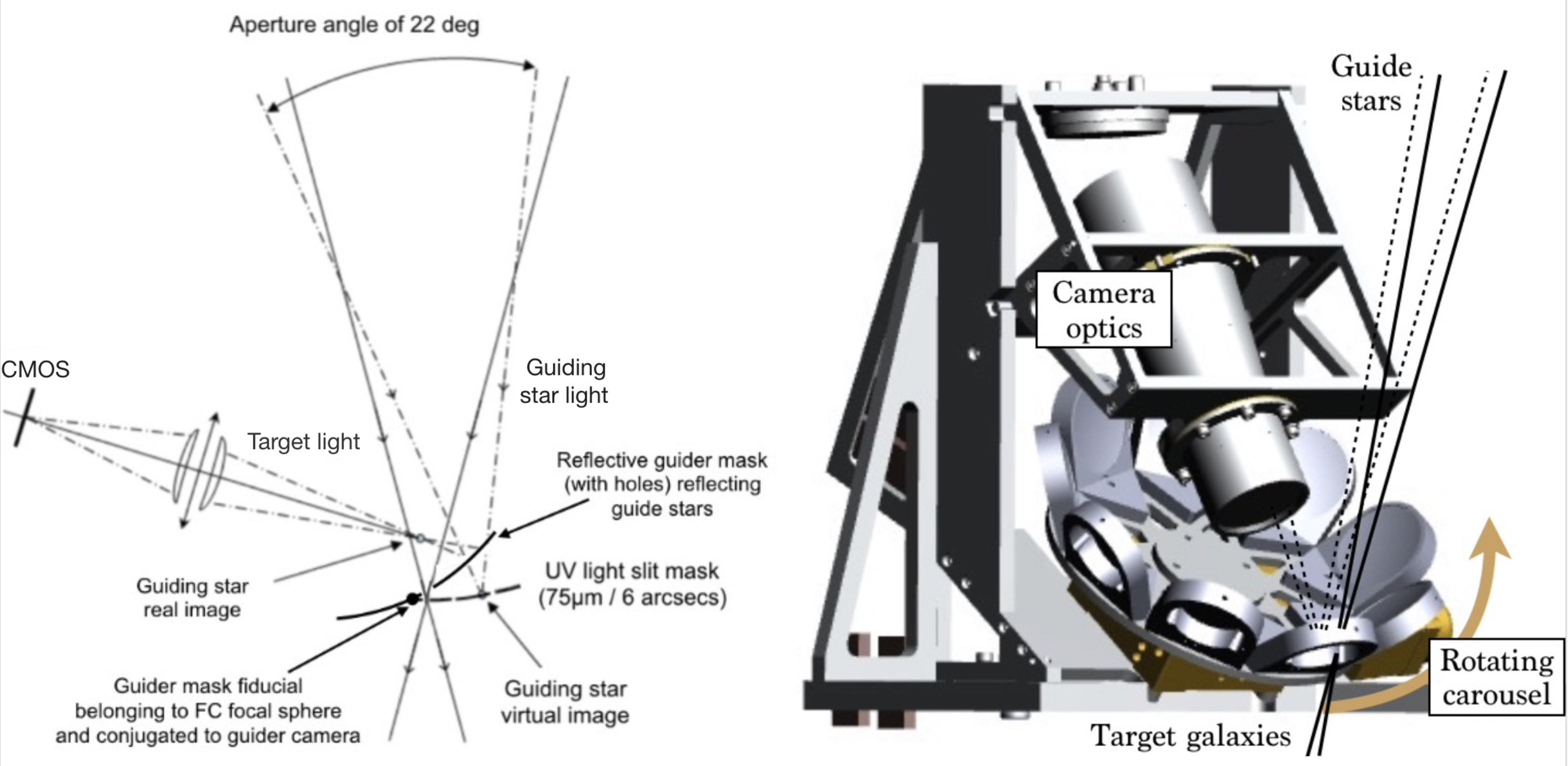}\\
\caption{\textbf{Left}: Mask-Guider-System (MGS) description scheme. The guider mask is a reflecting mirror,  just above the science maks,with holes at the target locations to let the light enter the spectrograph. The guiding specification is $1$ arcsec rms per axis for X and Y, but the hanging ground test on a bench simulating the complete flight chain have achieved a few tenth of arc-seconds. \cite{Montel2019}. The $3^{rd}$ guidance axis is provided by a rotating stage moving the complete payload guided to $1$ arcmin so that its contribution to the image is small. \textbf{Right}: Each mask unit, consisting of a guider mask attached to a science mask, is mounted on a rotating carousel and confocal  within $\pm50\mu m$. Dotted lines represent light from guide stars which are reflected, the solid lines are target galaxies, which pass through the slit mask. }
\label{MGS}
\end{figure*}

This section details the 5 steps required to align all subsystems inside the tank (sequentially: masks in the mask wheel $\rightarrow$ mask wheel in the guider system $\rightarrow$ guider system, spectrograph, FC, and detector on their bench) and place them at their nominal position in the medium bench mechanical reference (see Figure \ref{design} right).
\begin{enumerate}
\itemsep0em\setlength\itemsep{0.5em}
\item \textbf{Relative positioning of masks} to their nominal location with metrology feedback and iterative shimming.  The final measurements using a contactless \textit{STIL} distance chromatic sensor shows that all of the science masks are confocal within less than $\pm50$ microns.

\item \textbf{Positioning of the mask wheel} at a location that ensures the right platescale with Faro Arm feedback.

\item \textbf{Spectrograph and Focal corrector alignment} performed with Faro Arm feedback using a mechanical reference on the medium bench. From now on the spectograph and focal corrector assembly will be designated as \textit{the instrument} (Figure \ref{design}).

\item \textbf{Centering and focus of the guider.} The fiducial on the guider mask is conjugated with the guider camera (see Figure \ref{MGS}). Illuminated with visible light, this fiducial is positioned at the central pixel of the guider camera.

\item \textbf{Detector focusing} with feedback from full-field analysis of the resolution of the UV spectra of a Zn emission line lamp produced by a full-pupil diffuse illumination of the science and calibration masks. To infer the position of the focal plane, a tilted calibration mask that crosses the focus in both directions has been used initially and then refined directly on each science field image.  The final detector focus has been optimized to balance the image quality in the four science fields, so that the residual defocus is that generated by the scatter of the mask positioning ($\sim 25\mu m$ rms). 
\end{enumerate}

\subsubsection{Full instrument alignment}
This four-axis conjugation (two axis-centering operations and two tip-tilts) aims at aligning the instrument (spectrograph + FC) with the well aligned (steps 2 and 3 below) paraboloid and siderostat. After the instrument is aligned and centered on the paraboloid axis, a final through focus is to be achieved in flight to compensate for thermoelastic deformations. The ground conjugation is achieved in four steps: 
\begin{enumerate}

\itemsep0em\setlength\itemsep{0.5em}
\item \textbf{Centering the instrument} (tank) on its platform at the middle of its tip-tilt actuators, floor translation ranges and nominal elevation shims in order to have margin for the final alignment. We place a mirror on the upper part of the focal corrector assembly (called \textit{FC mirror} in the next steps).

\item \textbf{Aligning the parabola }(centering and alignment respectively at $\sim 1$ mm and $\sim 0.2$ degree precision) with the FC mirror by adding an orthogonal laser centered in the parabola hole. Center direct and reflected laser spots at the instrument top ensures the parabola has been roughly set with its axis close to that of the instrument.

\item \textbf{Finding the values of the sidereostat tip-tilt encoder position where it is orthogonal to the parabola axis (autocollimation).}  A movable $f/2.5$ visible light upwards source is placed near the parabola focal point  and adjusted along with the sidereostat orientation to minimize the coma of the return image. A byproduct of this operation is a good indication of the focal point location. 

\item \textbf{Final and accurate centering of the instrument.} The siderostat is set to autocollimation position (cf 3) and the mask is illuminated with visible light. The tip tilt ficus mechanism is adjusted to superimpose in the guider the direct and returned images of the illuminated mask.

\end{enumerate}

\subsubsection{Guider mapping and alignment}\label{sec223}

Such a balloon-borne multi-object spectrograph is at the limit of complexity of what can be calibrated at a launch base with extremely limited ground support equipment (GSE). It requires very demanding plate-scale verification ($\sim0.5\%$) and absolute X-Y in-flight positioning ($\sim 1''$) of the targets because their masks must be cut before the launch. Indeed, a major challenge of the FIREBall-2 project was to develop a $100\%$ self-consistent method that does not require any GSE for all of the pre-flight adjustments and calibration of a space facility. To this end, an invaluable asset of FIREBall-2 is its siderostat, which provides a full-pupil autocollimation capability \footnote{This is made possible by a good enough atmospheric transmission at 200 nm up to distances of a few tens meters at ground level.}.

The goal of the XY calibration is to calibrate the science mask centering relative to the guider, and determine on which exact pixels in the guider the guide stars should be placed to ensure that each target UV light is well centered into the 6-arcsecond slits ($\sim 80 \mu m$).
This must be achieved at arcsecond precision, which prevents the use of a collimator much smaller than the primary mirror because the residual aberrations would bias the calibration.

\begin{figure*}
\centering
\includegraphics[width=1\textwidth]{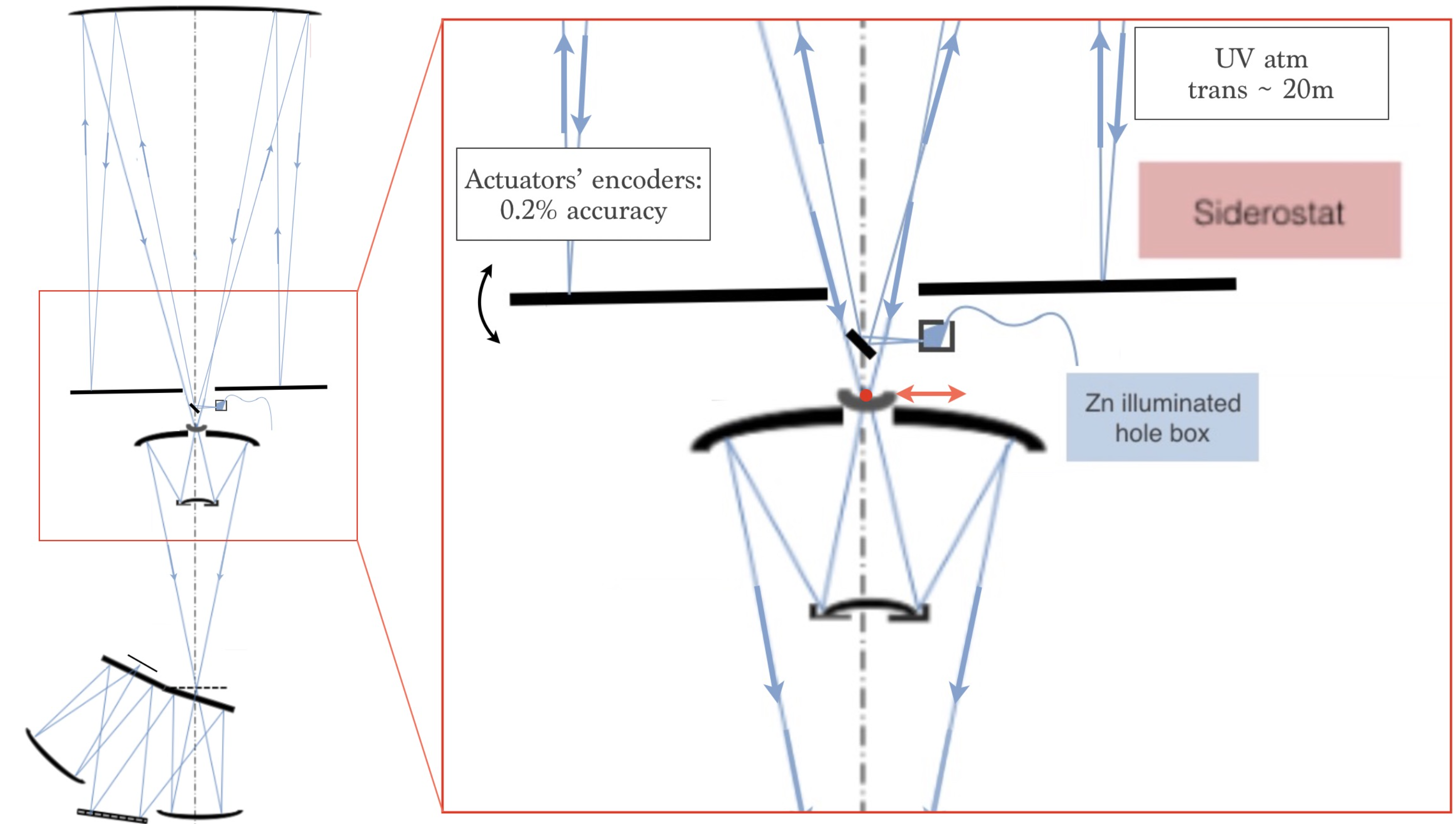}\\
\caption{\textbf{Left}: FIREBall-2 instrument in full-pupil autocollimation mode. \textbf{Right}: Zoom on the actuated part. Moving the siderostat actuators with high accuracy (black arrow) allows to move the image of the fiber (red dot on the siderostat focal plane) all over the detector plane while staying in full-pupil illumination, which is necessary to avoid sub-pupil effects that would improve artificially the point spread function (PSF) of the large optics.}
\label{autocoll}
\end{figure*}

To this end, the full-pupil autocollimation (see Figure \ref{autocoll}) mode has been used extensively in this calibration in order to avoid sub-pupil effects on image quality. The siderostat encoders are first calibrated against star images acquired with the guider at $\sim 0.2\%$ accuracy. Note that this calibration is independent of the guider distortion.
Then by putting a fiber illuminated by a Zn lamp (emission lines at $\lambda = $ $202.61371$ nm, $206.26604$ nm, $213.92635$ nm under vacuum) at the focal plane of the 1-m paraboloid (facing up) and by moving the siderostat, we are able to perform any desired angular move of the observed spot anywhere in the UV detector plane (without a mask) and the guider plane (for each mask), independently of the dispersion.

This capability is used to attach the pixel position of the guider stars to the targets position in the sky. A detailed description of this method is beyond the scope of this paper.

\subsubsection{Focusing}

The autocollimation mode has also been used to achieve and assess a 5-arcsecond image quality over the $0.6$-degree UV FOV using the full  $f/2.5$ aperture ratio to maximize precision.

Because only the guider images can be used to focus the instrument while in flight, the calibration procedure has provided the offset between the guide stars focus in visible light in the guider, and that of the targets in UV at the spectrograph detector. To this end, series of throughfocus have been performed at different positions in the field with the same setup as in \ref{sec223}.
In flight, the best UV focus is thus reached by setting the instrument to the calibrated defocus of the guide stars.

\subsection{Pre-flight performance}

\subsubsection{Instrument resolution}\label{subsubsec:resolution}
\textbf{Spectral resolution}\\

\begin{figure*}
\centering
\includegraphics[width=1\textwidth]{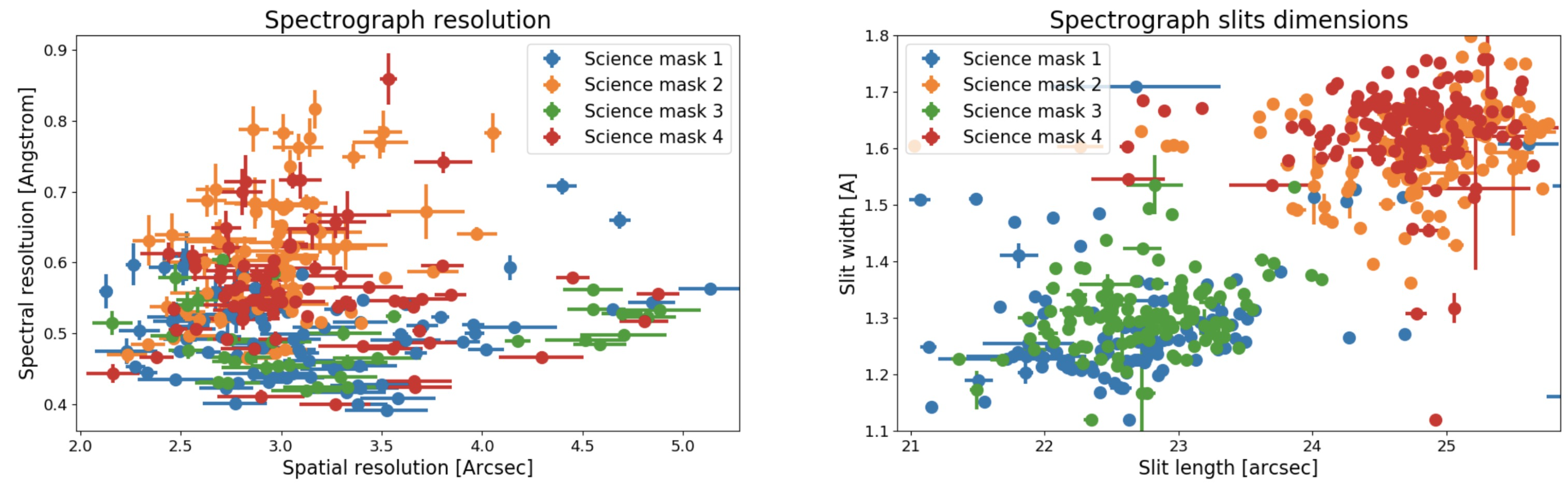}\\
\caption{\textbf{Left}: Spatial and spectral resolution FWHM of the spectrograph alone for a point source object. For diffuse objects the spectral resolution must be convolved by the $\sim1.5$ \AA{}  slit width. The resolution's scatter is mostly due to the shape of the masks. As the slits are located between the focal corrector and the spectrograph where the focal plane is spherical, the slits have been cut on a sphere. Some masks depart slightly from this nominal sphere which generates a slight defocus across the field of view.
\textbf{Right}: Slits' spatial and spectral dimensions for the four science masks. Two sets of masks are apparent, corresponding to a manufacturing with a different laser cutting setup.}
\label{slit dim}
\end{figure*}

The spectrograph spectral and angular PSF were measured before the instrument integration. This was performed using an holographic UV diffuser covering the whole UV field with an $f/2.5$ aperture and fed by a Zn emission line lamp. The diffuser was set near the focal surface at the payload entrance and was found to provide a uniform illumination to netter better than a factor 2.  It illuminates in turn each science mask so that the image at the detector shows well separated images of the three UV Zn lines over the whole field.
The detector was then shimmed iteratively to analyze the evolution of the PSF at three wavelength (\ref{sec223}) with the focus. After integration, the spectrograph resolution was re-evaluated for each science mask by applying simple de-convolution algorithms. This technique measures both the dimensions of the slits and the resolution of spectrograph. The results for each mask are shown in Figure \ref{slit dim}. 
For a point source object the spectral resolution of the spectrograph is found to be  $\sim 0.5$ \AA{} FWHM ($R\sim2800\lambda/d\lambda$).  
As the PSF at the science mask level is smaller than the slit's width, adding quadratically the angular resolution of the guider, the large optics and the focal corrector (see below) yields to an overall spectral resolution for a point source object of $R\sim2000\lambda/d\lambda$.

For diffuse objects, the spectral resolution is deteriorated by the spectral dimension of the slit. This leads to a $\lambda/d\lambda$ spectral resolution between $1300$ and $1600$ for the two sets of masks, allowing the spectral separation of components with a velocity difference larger than $200$ $km/s$.
The mean dispersion is found to be $\sim 608\pm 2$ $\mu$m/nm in accordance with the ZEMAX optical model, which predicts $\sim 606 \pm 0.5$ $\mu$m/nm \cite{mege2015}. \\

\textbf{Spatial resolution}\\
The resolution of each subsystem of the instrument has been evaluated independently using full-pupil illumination and applying a deconvolution algorithm.

The contribution to the spatial resolution of the guider, the large optics, the focal corrector, and the spectrograph are respectively $1.8''$, $1.8''$, $2.4''$, and $2.8''$ FWHM (see Figure \ref{Summary of FIREBall-2 sub-systems resolution, autocollimation resolution and in-flight prediction resolution}).

The convolution of the different optical sub-systems yields a FWHM of $\sim 4.5''$.  In parallel, end-to-end images have been taken with a $8$-arcsecond-diameter fiber at the parabola focal plane to illuminate the instrument in autocollimation (Figure \ref{autocoll}): the beam reflects off the large optics (twice on the paraboloid) and then goes through the total payload (focal corrector and spectrograph). After 2-D deconvolution of the fiber disk, the image at the detector level predicts an overall spatial resolution of $\sim 4.9''$ FWHM, which is a conservative measurement as it implies a double pass on the paraboloid and because the fiber was not yet optimally at the center of the FOV.
Due to mechanical deformations and in-flight defocus, we expect a small ground-to-flight image deterioration: the science masks' Z position are known with an accuracy of $\sim 50$ $\mu m$, which generates a $1''$ FWHM degradation (tested under \textit{ZEMAX}).

%
%

\begin{figure}[H]
\centering
\includegraphics[width=0.5\textwidth]{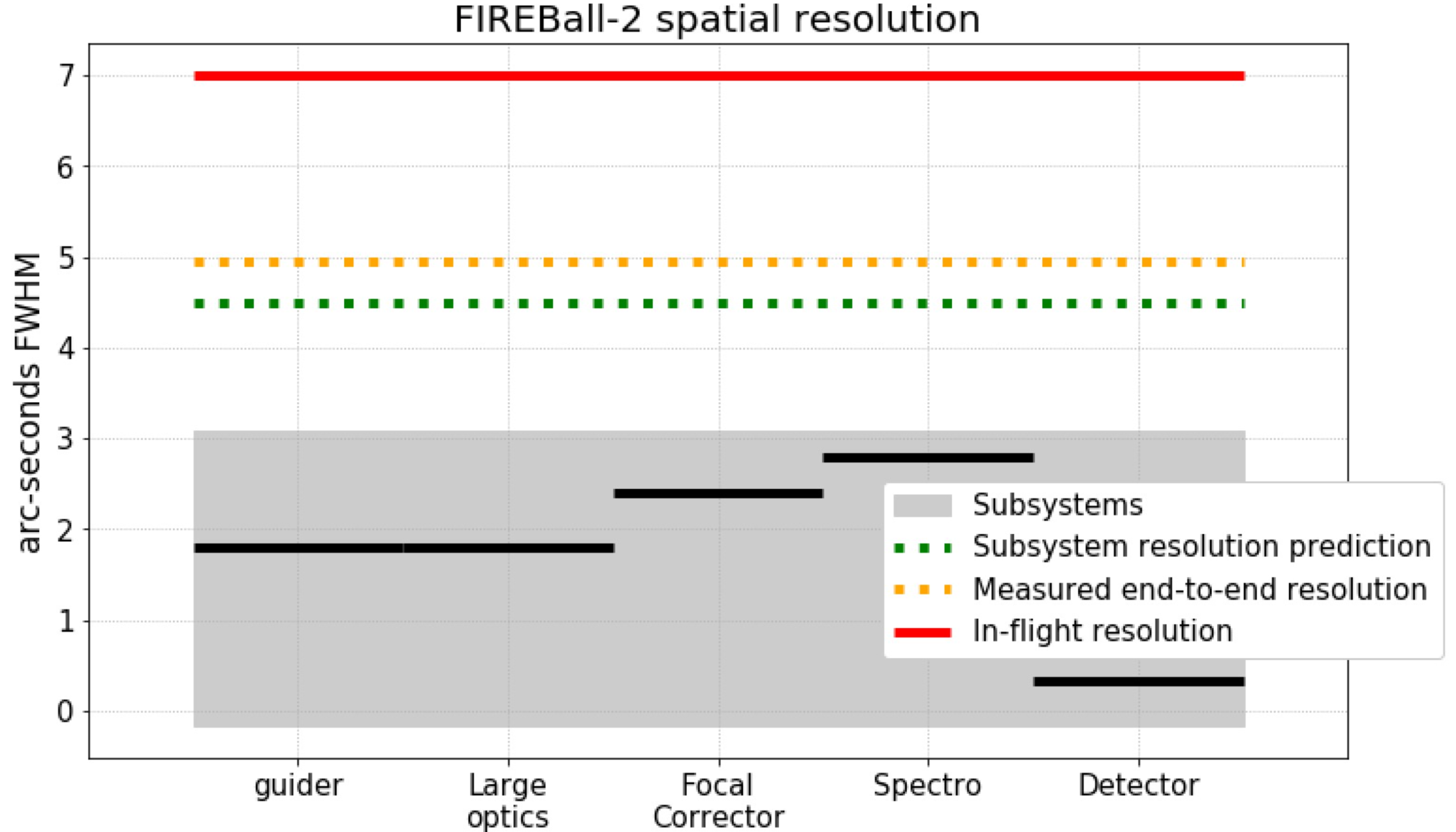}\\
    \caption{Contribution of the different subsystems to the FWHM spatial resolution in arcseconds (black lines). The degradation due to the pixels' size ($\sim1.1''$/pix) is negligible and taken into account in the resolution measurement performed on the flight-detector images.
    The conservative end-to-end measurement in autocollimation is plotted in red. The final prediction of the instrument in-flight resolution is plotted in orange and the flight value in red.} 
    \label{Summary of FIREBall-2 sub-systems resolution, autocollimation resolution and in-flight prediction resolution}
\end{figure}

\subsubsection{Throughput and optics' transmission}\label{sssec:num1}
A key optical component of the FIREBall-2 spectrograph design is the high-    throughput cost-effective, holographic $2400$ l/mm, $110$x$130$ mm$^2$ aspherized reflective grating used in the range $199$ - $213$ nm, near the $28^\circ$ deviation angle \cite{Quiret2014}.

The throughput of the overall instrument is measured by comparing input and output fluxes with a solar blind CsTe photocathode cell from ITT, which gives a $17\%$ efficiency for the instrument alone.

The total FIREBall-2 throughput is found to be $11\%$ (large optics: $65\%$, instrument: $17\%$).\footnote{Vignetting is included, detector quantum efficiency is not.} This is a factor of $6$ above what has been observed in flight for FIREBall-1 \cite{Tuttle2008,Milliard}. This has been achieved despite the more complex optics required to correct the larger field of view and double the angular resolution for this new generation of the FIREBall instrument. The throughput gain results  from the $55\%$ efficiency at $204$ nm of the FIREBall-2 grating (versus $32\%$) and from the large losses in the FIREBall-1 fibers. The additional gain of a factor $6$ on the detector quantum efficiency combined with the above and the improved guidance provides an order of magnitude gain in faint source detection efficiency to FIREBall-2.

\section{2018 Flight and data processing}\label{sec3}

\subsection{Flight}\label{sec31}
FIREBAll-2 was launched on a  $1.1$-million-cubic-meters balloon from the Fort Sumner Columbia Scientific Balloon Facility on September  $22^{\textrm{nd}}$  at 10:20 AM local MDT.  The flight sequence is detailed in \textit{Hamden et al. (2020)}. \cite{hamden2020}

After $\sim 3$ hours of ascent and $\sim 3$ hours at float altitude, the balloon began to lose altitude around sunset, most likely due to a hole in the balloon.  Because of this anomaly, the balloon dropped to its minimal mission requirement altitude after 45 minutes of science acquisition instead of the $>6$h required. Balloon altitude is important for science as between 30 and 40 km altitude, each km of altitude loss represents a reduction of transmission of $\sim5\%$.

In addition, weather constraints resulted in a launch the day before a full moon (moon $94\%$ percent full during the flight). The deshape of the balloon due to the balloon's deflation also impacted science data by concentrating a specular reflection of the moonlight over the balloon directly at the entrance of the payload. This boosted our background levels by $>1000$ times what was expected.
The resulting scattered light produced $\sim1$ event/pix/50s frame background, nearly an order of magnitude too high to efficiently perform photon-counting.\cite{Harpsoe2012}

Aside from these considerations, almost all subsystems performed nominally. Some flight anomalies are detailed in \textit{Hamden et al. (2020)} \cite{hamden2020} as well as the four fields observed. In order to focus on the instrument performance, we will only detail the reduction and analysis of the first field (DEEP 2:  RA=$253.06$, DEC=$34.97$, $\theta=20^\circ$) when the instrument was at the nominal altitude.

\subsection{DS9 quick-look and reduction pipeline}
In order to analyze flight data in real time, we built a python FIREBall dedicated DS9 extension similar to the DS9 \cite{Payne2003} Quick Look plugin (\url{https://people.lam.fr/picouet.vincent/pyds9plugin}).
Indeed, DS9 provides easy communication with external analysis tasks and is extensible via XPA. This extension was designed for real time processing in order to support in-flight decisions as well as assembly, integration and tests (AIT). 
During the flight it was mostly used for cosmic ray (CR) removal, background subtraction, help with object recognition, and stacking.

This extension proved its worth during the AIT and the flight, so the same DS9-oriented method was used to design the entire reduction pipeline to process data after the flight.
This extension is divided into three main parts, each of which fills a specific need:
\begin{description}
\item{\textbf{AIT}:} Supports assembly, integration, and test. This include throughfocus and through-slit visualization and analysis, XY calibration analysis, radial and encircled energy profiles, etc.
\item{\textbf{Flight reduction pipeline}:}  Contains every part of the flight data processing: Cosmic ray removal, over-scan correction, background subtraction, photon-counting processes, and image/object stacking.
\item{\textbf{Detector}:} Supports EMCCD characterization prior to and after the flight: gain computation from fluctuation analysis or histogram fitting, computation of the different noise sources ( clock-induced charges (CIC), read-out noise, etc.) and other tests: auto-correlation, column-line correlation, smearing analysis, etc.
\end{description}

\subsection{Data processing}

\subsubsection{Cosmic ray removal}\label{sec331}

Cosmic rays and other high-energy particles from outer space produce spurious features when hitting the detector. These random and unavoidable events generate spikes that are amplified in EMCCDs like any other photo-electron.
Each impact creates a “tail” in the direction of charge transfer due to deferred and overspill charge in the serial multiplication register.
The tail of the impact is highly dependent on the particle's energy. We note that each event profile on the detector has unique properties that might also depend on other characteristics, e.g., the integration time, the electron multiplication (EM) gain (and thus the degree of saturation in the high-gain register), the angle of the impact, etc. 

In order to identify and remove cosmic rays' peak and tail in the EMCCD images, we implemented in the DS9 extension a detection, identification, and removal algorithm. Detection is quite simple as any impact saturates at least one pixel on the device. 
The algorithm is based on a marching square code that creates isolevels isolating cosmic ray impacts. It incorporates a basic but efficient classification code in order to apply a more or less important mask depending on the particle energy. 

As we can detect up to 1000 cosmic rays in a 100-second exposure,  we had to optimize the speed performance of the detection/masking algorithm. After optimization, the process takes approximately 20 ms/impact on a $2.6$ GHz Intel Core $i7$ processor. 

We calculate a rate of $\sim2.4$ particles $\cdot$ cm$^{-2} \cdot$ sec$^{-1}$ in flight at 30-km altitude compared to $\sim0.05$ particles $\cdot$ cm$^{-2} \cdot$ sec$^{-1}$ on ground for the same CCD201 (13-$\mu$m pixel pitch) setup.
Between 26 and 33 km above the ground, we find a small increase of the impact rate with altitude of about $+1\%$  per kilometer.

These cosmic rays generate important signal-to-noise loss that affects about $0.05\%$ of the image per impact. On the 50-s exposure science frames, this represents $25\%$ of data that must be masked  in order to be fully conservative. 
This is equivalent to reducing the number of effective images by $25\%$, which prevents FIREBall-2 from capturing images with too long of an exposure time. 
In this respect, this constraint will be used in order to estimate optimum single-frame integration time, currently estimated to be  $\sim 30 - 80$s in order to maximize the signal-to-noise-ratio (SNR) (see section \ref{sec:snr}).

\subsubsection{Systematics contribution removal}
Before further processing, we need to remove systematic contributions in the EMCCD, especially account for the non-uniformity of the bias, and line-to-line and column-to-column variations.

The line-to-line variations seem to come from remaining charges in the serial amplifier or poor amplification register clearance that do not always empty it to the same level. Because of a higher charge transfer inefficiency for these highly saturated events, some cosmic ray tails can continue on the left over-scan region\footnote{ for extremely energetic events $(\sim 0.1\%)$ it also appears that the charges of the cosmic rays are not cleared and then continue over several lines from the left over-scan to the right.}. Because of this, to estimate the bias level on each line, we take the median value of the left part of the right  over-scan region (between column 2200 and 2400), just after charges have been cleared out from the serial register. This prevents including in the bias any background contribution in the spectral direction.

We then correct the column to column variation by subtracting for each column its median value. Then the corrected value of the pixel in row \textit{i} and column \textit{j} is:
$$p^{corr}_{i,j} = p_{i,j} - \widetilde{p_{2200<\cdot<2400, j}} -  \widetilde{p_{i, 0<\cdot<2069}} $$

The dot in the index is a placeholder to indicate whether the
mean was taken over rows or columns. We then use this notation $\widetilde{p_{n_1<\cdot<n_2, j}}$ for the \emph{median} value of the $j^{th}$ line between column $n_1$ and column  $n_2$.

Note that the subtraction of the column medians in flight images with background does not remove the potential detections, as they overlap only a few lines and are completely washed out in the subtracted medians. In addition, the signal coming from the sky and reflections in the instrument, and even more its series of vertical medians, has only low spatial frequency components, which do not affect our potential detections expected over only a few pixels.

\subsubsection{Background subtraction}

The moon's light reflected on the balloon (see Section \ref{sec31}) creates high sky background count rates of $2\times10^6$ Line Units (LU) \footnote{Line Units : $n_{photon} \cdot s^{-1} \cdot cm^{-2} \cdot sr^{-1}$} for diffuse reflection, which is more than $2$ orders of magnitude greater than what was expected.

In addition, at very specific azimuth angle differences between the instrument, the moon, and the balloon, reflective parts in the instrument are illuminated, increasing the parasitic light  10 times higher than the diffuse reflection.

Since the detector is not solar blind and we do not see any correlation between the background intensity and the altitude over a factor 5 in atmospheric transmission, we infer that the background comes mostly from visible light leaks. Post-flight accurate scattering analysis shows that the incoming light from the balloon was not following the nominal optical path but was reflecting from different mechanical components inside the tank.

As the background changes quickly in shape and intensity between exposures, and since the CR masking modifies significantly which exposures are retained in the summation for each pixel, the mean background value varies from pixel to pixel. Therefore the background must be estimated and removed individually on each cosmic ray free image.

\begin{figure*}
\centering
\includegraphics[width=.9\textwidth]{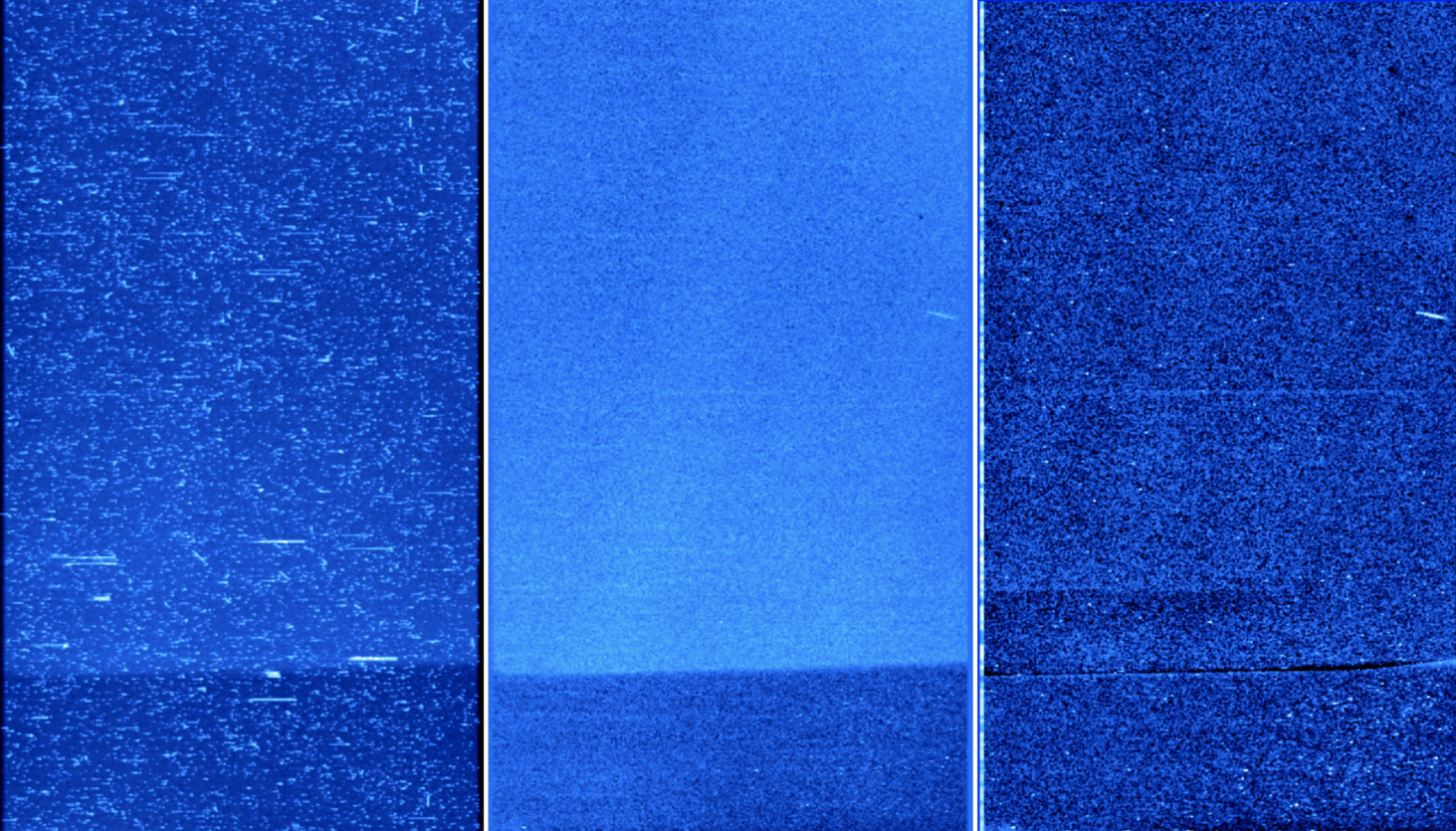}\\
\caption{Field 2 summed images. \textbf{Left}: The summed image before cosmic ray removal. \textbf{Center:} The summed image after cosmic ray removal. \textbf{Right:} The summed image after cosmic ray removal and subtracting the out-of-band background scattered light. The sharp drop-off in scattered light on the bottom fifth of the image makes the subtraction difficult, yielding a residual along the edge of the scattered region. All images have been smoothed with a Gaussian kernel with sigma of 1.5 and a radius of 3 pixels. The emission from the brightest star, Bright Star 1, is visible in the last two images as a horizontal line roughly in the middle \cite{hamden2020}.}
\label{reduc}
\end{figure*}

Besides this, there is a curved and net boundary in the image background (see Figure \ref{reduc}). This break, most probably due to a component shading the bottom part of the detector, makes automatic background subtraction difficult. In order to handle it, we divide the image in two curved sub-images and perform background subtraction on each independently.

The average background level is computed iteratively within $40$-pixel image bins by clipping out pixels at $3\sigma$ from the average (maximum 10 iterations) and taking the median value. 

The low-resolution background image is then convolved by a $3 \times 3$ 2D median filter and is then re-sized to the original data size using spline interpolation. 

Because it is negligible compared to the high spurious flight background, we do not need to correct for dark current ($0.08$ e$^-$/h/pix on ground).

\subsubsection{Stacking and hot pixel removal}
The images are stacked by taking the mean value at each pixel position. Indeed as the nominal flux level we want to detect are lower than $0.5$ e$^-$/pixel, taking the median would throw away important detections.

However, some hot pixels show up after stacking the images. Because they have been identified on ground dark images, and they only spread onto 3-4 pixels in the charge transfer direction, we masked every 4 pixels following a hot pixel identified on ground and interpolate the values of the pixels inside the mask by applying a  2D $\sigma=1$ pixel gaussian filter.

\subsubsection{Flux map generation}\label{sec335}

\begin{figure*}
\centering
\includegraphics[width=.85\textwidth]{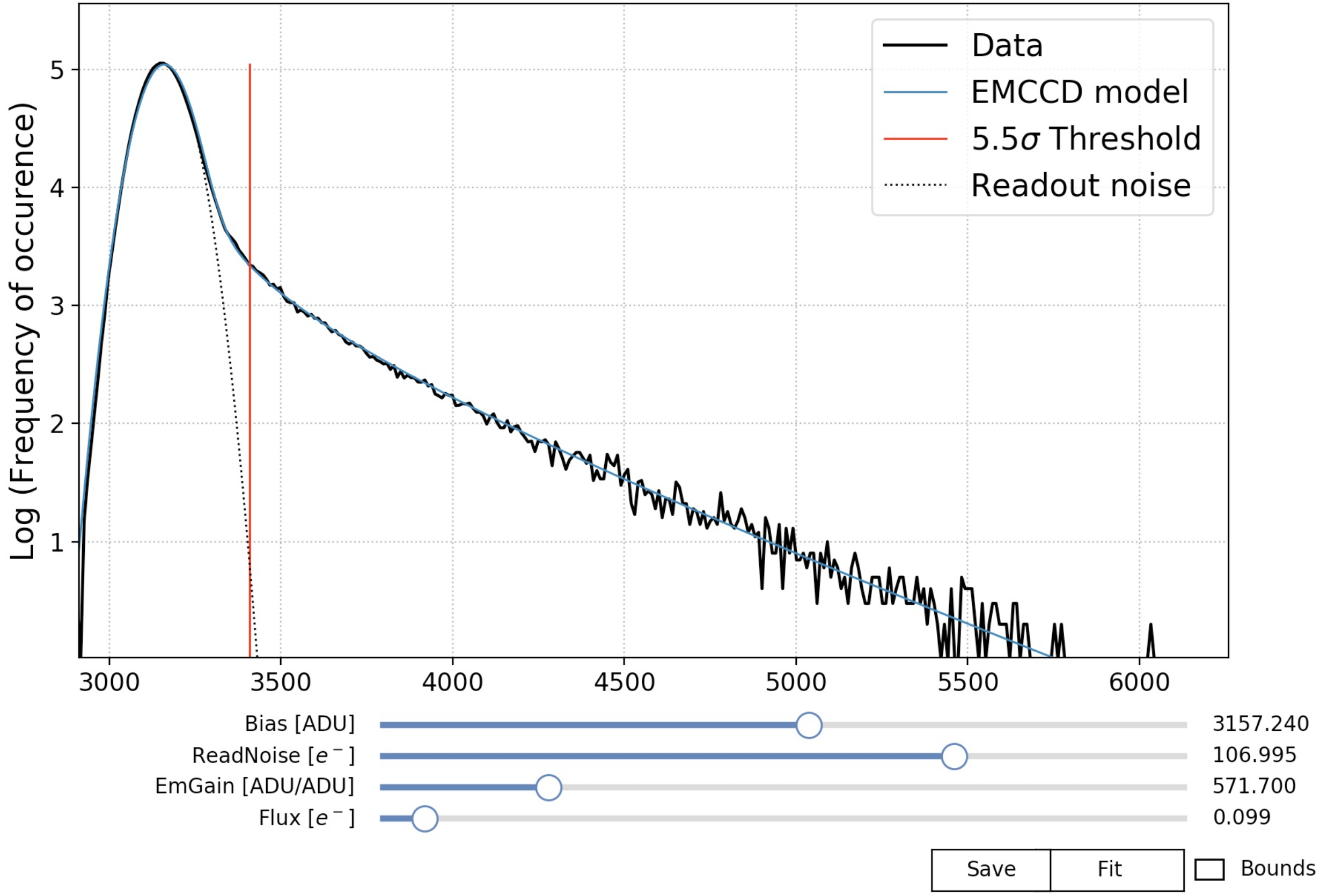}\\
\caption{Unique electron spectrum. This fitting algorithm from the FIREBall DS9 extension allows fitting and recovering EMCCD parameters in the absence of smearing. The red line represents the photo-counting threshold used to mitigate the excess noise factor due to amplification. The dotted black line represents the readout noise. In photon-counting mode, this noise is lowered to the part above the $5.5\sigma$ threshold. As explained in Section \ref{sec335}, the histogram shape is affected by the smearing that acts exactly as if the amplification gain was lower ($1400$e$^-$/e$^-\rightarrow 600$e$^-$/e$^-$).}
\label{hist}
\end{figure*}

\textbf{Assessment of the amplification gain}\\
The gain is needed to convert the measured ADU levels to photoelectrons, but careful calibration is required because of its dependency on detector temperature ($\sim 10\%$  per degree at $T  \sim -110^\circ $C and $G_{amp} \sim 1000$ e$^-$ / e$^-$).
Three methods are usually used to measure the amplification gain of the detector:
\begin{itemize}

    \item \emph{Unique electron spectrum} from a flat field image having a low illumination level of $< 1$ e$^-$/pix. Looking at the histogram of ADU values with logarithmic counts, the bias corresponds to the mode (most frequent ADU value), the readout noise to the width of the parabola around the bias, obtained through a quadratic fit, and the gain to the slope in the amplified part (highest values) \cite{Harpsoe2012}. See Figure \ref{hist}.
    \item \emph{Fluctuation analysis }derived from the slope of the variance/mean ADU diagram without any correction. This gain includes the amplification process and the conversion gain \cite{Hirsch2013}.
    \item \emph{Direct measurement} by dividing two images with and without gain. The input flux must be as stable as possible and the exposure times must be different so that we stay far from the dark level for the non-amplified image and far from the saturation level for the amplified image.   
\end{itemize}
Because the two first measurements can be significantly impacted if smearing appears in the CCD or in the amplification registers, the gain in flight has been post-calibrated on ground using the third method at the flight temperature. Indeed, during the flight, the smearing exponential length of $\sim 1$ pixel drastically affects the amplification gain derivation from two first methods ($\sim1400$ e$^-$/e$^-$ $\rightarrow \sim650$e$^-$/e$^-$ ). Using method 3, we retrieve a flight amplification gain of 1370.
This gain was corroborated by a complete electron spectrum model, including a toy linear model of the smearing effect, that was tested at several illumination levels.\\

\textbf{Fluxes outside atmosphere}\\

We can then compute the flux in the co-add image:

$$F_{e^-/pix} = \frac{pix_{ADU}\times G_{conv}}{G_{amp}\times t_{exp}}$$

with $pix_{ADU}$ being each pixel value in ADU, $G_{conv}$ the conversion gain ($1.8$ e$^-$/pix), $G_{amp}$ the amplification gain ($\sim1370$ e$^-$/e$^-$) and $t_{exp}$ (50s) the exposure time.

In order to retrieve the flux in photons per Angstrom at the detector level:
$$F_{\gamma^-/\AA{}}^{Det} = \frac{F_{e^-/pix} \times dispersion }{QE}$$

where $QE\sim50\%$  is the quantum efficiency of the detector. The dispersion of the instrument is $\sim 4.66$ pix/\AA{}. 

Then, above the atmosphere the flux is:

$$F_{\gamma/\AA/cm^2} = \frac{F_{\gamma^-/\AA}^{Det} }{A_{telescope} \times t_{atm} \times t_{FB}}$$

where $A_{telescope}=7854$ cm$^2$ is the collecting area of the telescope, $t_{atm}\sim 33\%$ is the transmission of the atmosphere at the telescope average altitude, and $t_{FB}$ ($\sim13\%$ including large optics and vignetting) is the throughput of the FIREBall instrument.

\section{In-flight performance}\label{sec4}

\subsection{Detection and proof of performance}\label{sec41}

The ground-based calibration for absolute pointing and XY positioning described in Section \ref{sec3} assumes a stability better than 1 arcsec of the guider-to-mask relative positioning until the end of the flight, in an environment $40^\circ$C colder than during the calibration. 

The continua of the three UV-bright \textit{GALEX} stars in the science field were detected, with a SNR $\sim18$ for the brightest one ($M_{FUV}$=17.8). 
This proves that the targets were positioned close to their nominal position, one of the toughest challenges of FIREBall-2. Flight data do not allow us to access the residual decentering between the source and the exact slit center. Photometric measurements are performed in Section \ref{sec45}.

\subsection{Guidance performance}
In addition to its impact on the throughput and flight duration shown in the previous section, the loss of altitude also impacted the balloon's (and consequentially the gondola's) motion because of the varying wind direction with respect to altitude. This intensified the different gondola excitation modes, requiring better damping and causing a decrease in the pointing performance. Despite that, the pointing performance stayed well within the technical specification \cite{hamden2020}.

The in-flight raw guider error signal is found to be $0.7''$ RMS  per axis, an upper limit to the guidance error, as it includes the guider signal noise and the field rotation guiding error of $\sim0.3''$ RMS. The latter number derives from a $1.5$ arcmin field rotation 1D guidance error, and a typical distance guiding stars-targets of 12 arcminutes\cite{Montel2019}.

\begin{figure*}
\centering
	\includegraphics[width=1\textwidth]{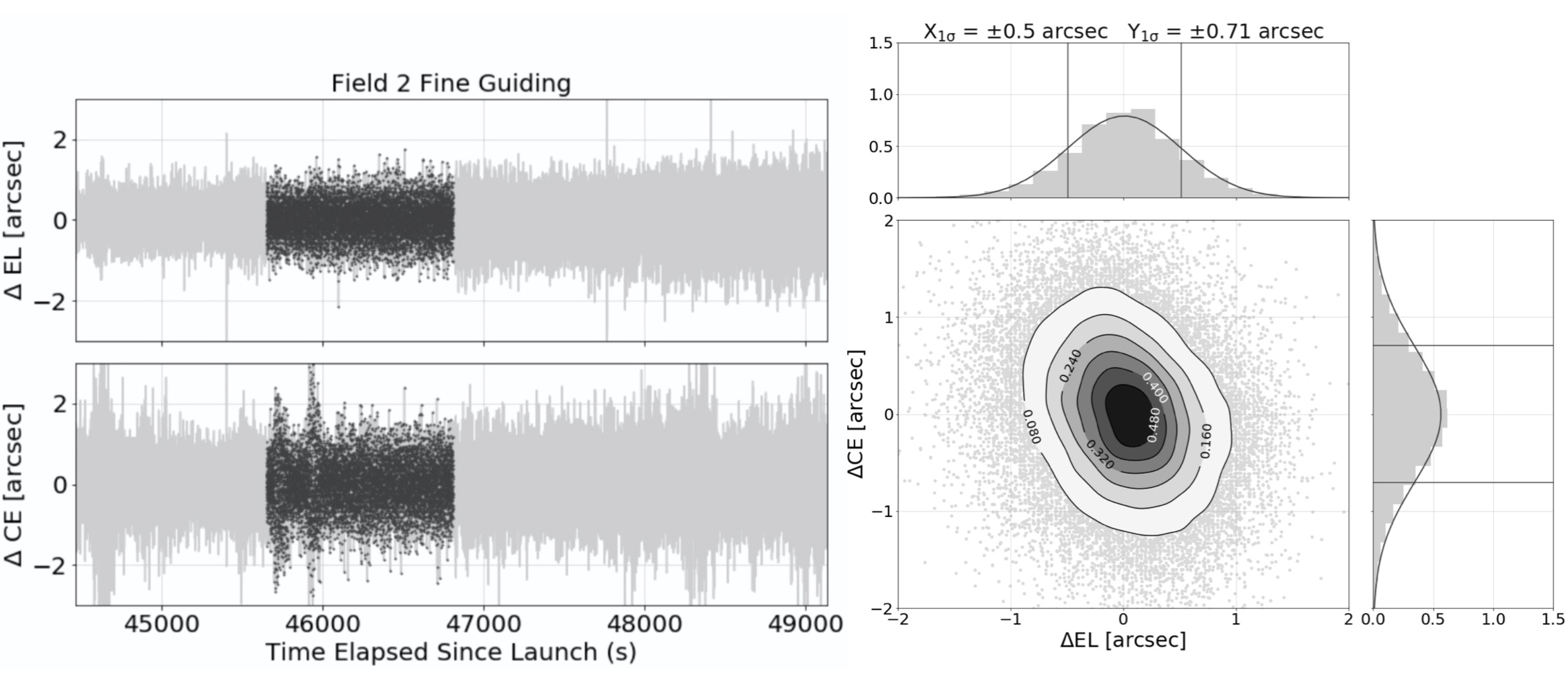}\\
\label{pointing}
\caption{\textbf{Left} Evolution of in-flight fine guiding performance. \textbf{Right} Summary of the 2-axis (elevation and cross-elevation) fine guidance errors in closed loop on Field 2 during the $\sim1$-hour observation. Data corresponding to ballast drops has been excluded. The 1D probability density histograms are fit with a 1D gaussian, which we use to measure the $1\sigma$ errors in the 1D distribution (vertical lines). The 2D distribution in the center is interpolated with a gaussian kernel density estimation. The $1\sigma$ contour here corresponds to 0.39 arcsec$^{-2}$. FIREBall-2 achieves a ($1\sigma$) pointing of $\pm 0.5$ arcsec in elevation and $\pm 0.71$ arcsecond in cross-elevation (\textit{Melso et al, in prep.}) .}
 
\end{figure*}

\subsection{Image quality}

An in-flight spatial resolution of $7''$ has been derived from a gaussian fit along the spatial direction of the \textit{GALEX} brightest star continuum in the first field. We are currently working at LAM on improving the focus of the spectrograph in order to obtain an in-flight $5''$ FWHM  angular resolution all over the field for the upcoming flight in 2021. This is of course a very important gain in resolution and it would also decrease all the sources of noise in counts/PSF/hour as the PSF gets sharper (this is taken into account in Table \ref{blabla} and detailed in Section \ref{sec:snr}).

\subsection{Detector performance}
FIREBall-2 incorporates a state-of-the-art $2$Kx$1$K, $13$-$\mu m$-pixels electron-multiplying CCD 201 from \textit{Teledyne-e2v} processed at JPL to increase its UV sensitivity with a  delta-doping procedure, which yields a $100\%$ internal quantum efficiency device limited only by the reflection losses, and a multi-layer antireflection coating to further enhance external QE\cite{Nikzad}. The stochastic nature of the amplification process in EMCCDs generates an Excess Noise Factor (ENF=$\sqrt{2}$ at high gain) that can be removed under certain conditions ($Flux<<1$e$^-$/pix/exp and $G_{amp}>10\times RN$) by applying a thresholding process known as photon-counting \cite{Harpsoe2012}. The different characteristics of the detector are summarized in Table \ref{Flight UV EMCCD characteristics: w17d13}.
These state-of-the-art devices require special parameter tuning in order to take advantage of their performance, as well as specific data reduction such as cosmic ray removal or photon-counting processing.

The present model of FIREBall-2 detector has been calibrated on the ground and shows a $55\%$ peak efficiency in the nominal UV instrument bandpass and $>50\%$ in the entire bandpass.  This value is compared to flight data  in Section \ref{sec45}. The detector performance is detailed further in \textit{Kyne et al. (2020)}\cite{kyne2020deltadoped}.

Relevant to the present analysis, the detector shows sub-optimal charge transfer efficiency (CTE) due to the unusually low temperature  selected to minimize the dark current. We think this can be improved by further optimization of the controller/parameter tradeoff for the next flight. 

At flight temperature, this average CTE induces a charge smearing that yields a performance similar to an electron multiplying (EM) gain three times lower. This increases the noise contribution and causes event loss when attempting to perform photon-counting thresholding. This effect combined with the high background prevented us from using photon-counting.  Smearing inversion algorithms can be applied, but there is an intrinsic limitation as these inversions always generate noise amplification \cite{kyne2020deltadoped}.

\begin{table}[H]
\centering 
\renewcommand{\arraystretch}{1.4}
\caption{Flight UV EMCCD characteristics. $^\star$: Parameters measured on ground as they require extremely low illumination levels not obtained during the flight. The CIC we give here is a lower limit as we do not take into account horizontal CIC.}
\label{Flight UV EMCCD characteristics: w17d13}
\begin{tabular*}{\linewidth}{@{\extracolsep{\fill}}p{0.4\linewidth}p{0.22\linewidth}p{0.27\linewidth}@{}}
\hline
\textbf{Parameter} & \textbf{Unit}  & \textbf{Value}  \\ 
\hhline{===}
\textbf{Device's temperature} & T(C$^\circ$) & $-110$ \\ 
\textbf{Number of amplification pixels} &  &  $604$ \\ 
\textbf{Amplification} & ADU/e$^-$ & $1400\pm30$ \\ 
\textbf{Bias} & ADU  &  $1.4\times 10^3$\\ 
\textbf{Read noise (no photon-counting)} & $e^-$ &  $107\pm1$\\ 
\textbf{CIC$^\star$} & e$^-$/pix/exp &  $0.005\pm0.002$\\
\textbf{Dark current$^\star$} & e$^-$/pix/h & $0.08\pm0.03$ \\ 
\textbf{Smearing exponential length} & pix &  $0.7\pm0.1$ \\ 
\textbf{Exposure time} & seconds & $50$ \\
\hline 
\end{tabular*} \\
\end{table}

\subsection{2018 flight sensitivity}\label{sec45}

The total efficiency was computed based on the flux measured on Bright Star 1 using a fitting-based photometric algorithm. 
 The measured continuum was found to be $30\%$ lower than expected from the ground-based calibration and atmospheric model, which is well within statistical fluctuations derived from applying the same photometric algorithm at random locations. The main contributors to this uncertainty are: target centering, photometric measurement, atmospheric transmission model, focus, instrument throughput and detector calibration. In addition, the photometric measurement on BS1 shows a decreasing flux during the observation, which result probably from defocus or target decentering rather than atmospheric model bias.\\
No emission line has been detected at a statistically significant level, neither by searching at the predicted individual emission line locations, nor by stacking the target galaxies or a subset of the brightest ones.
In order to measure the sensitivity limit for the total Ly$\alpha$ emission line of a target, we have performed profile fitting of a 2D gaussian matched to the PSF at random locations in the image. We avoided  areas where target signal can be expected. The derived $5\sigma$ sensitivity limit for an unresolved $z\sim0.7$ source for the available 30-min observation is $\sim1.5 \times 10^{43}$erg/sec.
This limit is found consistent with a rather simple but quite complete analytical calculation of the total sensitivity of the instrument. As can be seen in Figure 7 of \textit{Augustin et al. (2019)} \cite{Augustin2019} FIREBall-2 is presently lacking a factor of 3 in sensitivity to constrain the models at $z\sim0.7$. 

For the diffuse CGM Ly$\alpha$ emission, the instrument-limiting flux $\sim7''$ away from a single object center, using one resolution element on each side of the central one, translates to a surface brightness detection limit of $~1.7\times10^{-16}$ergs/cm$^2$/s/arcsec$^2$ (721,000 LU).

Limiting flux performance is, of course, lower than what was expected without the spurious background but can be used as a starting point for evaluating the instrument's ultimate sensitivity.


\section{Ultimate sensitivity}

The operational success of the 2018 flight and clear paths towards correcting the issues that reduced the sensitivity motivated \textit{CNES} and \textit{NASA} to fund the project for at least one more flight.

\subsection{Major improvements for 2021 flight}\label{sec51}
Table \ref{blabla} describes the different sources of noise as observed in 2018 and expected in 2021. It shows that the major offender to the sensitivity is the high sky level, mostly due to the moon-scattered light, as discussed in Section \ref{sec41}. Because of the very short turnover wind period, launch opportunities are scarce and can occur very close to the full moon, as it happened in 2018. A launch far from full moon phase will always be preferable, but stray light mitigation has to be sufficient so that even in the worst \textit{scenario}, sky noise would not impact FIREBall-2's performance \cite{Hoadley}.

Our simple SNR model shows that to achieve this goal, it is necessary to cut stray light to the UV detector by a factor larger than 200. In this case, the sky background will be less than $3$ counts/PSF/hour. Refinements of the stray light/off-axis reflections analysis are on-going but it is already clear that three baffling stages need to be set up: a gondola top \textit{cap} combined with a lower baffle will prevent direct view of the balloon from the spectrograph entrance (direct view of the moon was already baffled). It will be aided by an improved system of baffles internal to the tank.
A more stringent moon-angular-distance limit and extensive testing of the baffling will ensure in-specs background in the worst expected conditions. 

We experienced some \textit{additional background noise} which we distinguish from the \textit{detector dark noise} (Table \ref{blabla}). 
Indeed, during the flight we observed a signal, when the shutter is closed, an order of magnitude higher than the ground-based dark at the same temperature. Even though it is not totally understood, we think it is not an increased detector dark but rather a consequence of either Cerenkov emission coming mostly from the back of the FF2 mirror, a guider camera leak or (less likely) a shutter failure that would have let $\sim1$cm$^2$ of light  through. We have decided to reduce any possible Cerenkov contribution by masking all the mirrors' backs and are currently addressing the other possible causes. We have also added more comprehensive baffling around the immediate detector environment. It is important to stress that the level of \textit{additional background} (shutter closed) might come from other unknown causes as it is the very first flight of these-state-of-the-art devices (large number of very low energy cosmic rays or longer than estimated smeared cosmic ray tails).

In addition, a new detector system will be used for the next flight. It will equipped with a new version of the detector controller (\nuvu  v3), developed specifically by \nuvu Cameras  for these EMCCD detectors.\cite{kyne2020deltadoped} The expected lower read noise provided by this controller combined with a slightly higher temperature should make the photon-counting algorithm more efficient. Data acquired with this new setup will allow further optimization of  the temperature/smearing/gain  parameters in order to increase the SNR.

The improvements in the spectrograph alignment and guidance algorithm also promise an image quality increase from $7''$ to $\sim5-6''$, which improves our chances of disentangling CGM signal from the galaxy signal and beginning to spatially-resolve CGM emission.

\subsection{2021 flight predicted sensitivity}\label{sec:snr}

\begin{table*}[t]
\centering 
\renewcommand{\arraystretch}{1.4}
\begin{tabular}{@{}llllll@{}}
\hline
\textbf{\begin{tabular}[c]{@{}l@{}} 2018 Parameter\\ Units = cts/PSF/Field (2 hours)\end{tabular}}  & \textbf{\begin{tabular}[c]{@{}l@{}} 2018 Flight\\ Analog mode\end{tabular}}  & \textbf{\begin{tabular}[c]{@{}l@{}} 2021+ Flight\\ Analog\end{tabular}} & \textbf{\begin{tabular}[c]{@{}l@{}} 2021+ Flight\\ Counting\end{tabular}}  & \textbf{Comment}                      \\
\hhline{======}

\textbf{Read-out noise (50s images)}            & 6.4        & 5       & 0.8       &    \nuvu  v3  and $ \gamma$-counting \\
\textbf{Clock induced noise}          & 8.3                        & 6.5       & 4.6       &   Only sharper PSF  \\
\textbf{Detector dark noise}          & 3.9                        & 3.0       & 2.1     &   Measured in lab  \\
\textbf{Additional background noise}  & 21.8                   &    5.4     &      3.8   &     Shutter close, Cerenkov? \\
\textbf{Sky noise}                              & 90& 5.8       & 3.4       &  See Better baffling see 4.1\\
\textbf{Shot noise}          &31.3  &     13.3  &  7.1  &  For a $5\sigma$ object\\
\hline
\textbf{Total noise}                            & $\sim$98.6 & $\sim$17.8 & $\sim$10.1 & $\sqrt{\sum Noise^2}$\\
\hline                 
\end{tabular}
\caption{FIREBall-2 noise [cts/PSF/Field (2 hours)] contribution using one resolution element on each side of the object. Aside from the read noise, all noise analog values account for the $\sqrt{2}$ excess noise factor coming from the stochastic nature of the CCD amplification. In photon-counting mode, this factor is close to 1. }
\label{blabla}
\end{table*}

In order to analyze the advantage of using photon-counting, we compare here two \textit{scenarios} for the next flight: with [\textit{Counting}] or without [\textit{Analog}] thresholding photon-counting process. We assume a 200-factor sky reduction and that the \textit{additional background level} (as defined in Section \ref{sec51}) is mitigated by a factor of 10. To compute the SNR, we use one resolution element on each side of the central one for 144 exposures, amounting to 2 hours observing time\footnote{The effect of cosmic ray mentioned in Section \ref{sec331} is taken into account. It is important to remind that when stacking over a long period of time, the SNR does not increase as fast as $\sqrt{t}$. We accounted for that with the same $1.4$ factor observed between $50$sec and $30$min.}.

The noise contributions for these two cases are reported in Table \ref{blabla} so that they can be compared with 2018 values. 

In both \textit{scenarios}, FIREBall-2 appears to be photon starved: the biggest offender (highest noise contribution) is the shot noise. In the photon-counting \textit{scenario}, stacking 10 targets with similar galaxy properties/morphologies would decrease the shot noise to a level comparable to the other noise sources. This  would make the major noise contributors (CIC, sky, background, shot noise) very similar, prohibiting any further gain. 

As expected in photon-counting mode, the read noise contribution becomes much smaller and part of the CIC noise disappears as the serial CIC is only amplified by an average gain of 200 instead of 1400 e$^-$/e$^-$. The Noise was computed assuming $30\%$ count loss and only $85\%$ read noise reduction because of smearing when applying the photon-counting thresholding process. These values are representative of the expected improvement of the charge transfer efficiency (CTE) with respect to the 2018 values. 

The fact that, in both \textit{scenarios}, the detector dark noise is not dominant might allow for a higher detector temperature, which would improve its CTE and facilitate photon-counting. 

For these two 2021 cases, the respective limiting $5\sigma$ total Ly$\alpha$ emissions are $\sim4.3\times10^{41}$ and $5.3\times10^{41}$erg/sec for a $\sim5'' $ unresolved $z\sim0.7$ target.
For the diffuse CGM Ly$\alpha$ emission limit and using the same definition as in Section \ref{sec4} (with $5''$), the surface brightness detection limits are respectively  $~1.4\times10^{-17}$ and $1.8\times10^{-17}$ ergs/cm$^2$/s/arcsec$^2$ (61,000 LU and 75,000 LU).
As we can see, these limiting fluxes for the two \textit{scenarios} are very close despite very different total noise. This is due to the fact that the high photon-counting threshold used to mitigate the read noise and the serial CIC, coupled with smearing, generates event loss and therefore lowers the total efficiency by $\sim30\%$.

In order to compare FIREBall-2's performance with that of the first generation of the instrument, we gathered in Table \ref{Improvements} the characteristics and performance of FIREBall-1 and 2 for their respective flights. In Table \ref{Improvements}, the sensitivity limit is computed for 2 hours acquisition on an extended $30''$ object such as a bright Quasi Stellar Object QSO.

\begin{table*}[t]
\centering 
\renewcommand{\arraystretch}{1.4}
\begin{tabular}{@{}llllll@{}}
\hline
\textbf{Parameter}                           & \textbf{\begin{tabular}[c]{@{}l@{}}FIREball-1\\ 2009 Flight\end{tabular}} & \textbf{\begin{tabular}[c]{@{}l@{}}FIREball-2\\ 2018 Flight\end{tabular}} & \textbf{\begin{tabular}[c]{@{}l@{}}FIREball-2\\ 2021+ Flight\end{tabular}} & \textbf{$R^\alpha$} & \textbf{Rationale}                      \\
\hhline{======}

\textbf{Detector QE}      & 0.07    & 0.55     & 0.55    & 8         & MCP $\rightarrow$ EMCCD   \\
\textbf{Effective QE}      & 0.07    & 0.41     & 0.37    & 5.3        &   CR impact, $\gamma$-counting loss \\
\textbf{Instrument efficiency}      & 0.0036              & 0.13      & 0.13     &    36      & Fiber $\rightarrow$ slit, grating, misc \\
\textbf{Atm transmission}             & 0.4                & 0.4                                                                & 0.5                                                                       &      1.25     & Higher altitude                         \\
\textbf{Net efficiency$^\beta$}                      &        0.0001      &            0.021         & 0.024 &      238    &       \\
\textbf{Spatial resolution FWHM}   & $12''$                 & $7''$                                                                       & $5''$                                                                   &       2.4    & Resolve CGM regions                     \\
\hline
\textbf{Total noise [cts/PSF/hr]}   & $\sim$0.7                 & $\sim$67  & $\sim$ 5 &     0.15      & 2021 optimistic \textit{scenario}                     \\

\hline

\textbf{Effective FOV (arcmin$^2$)}          & 8                   & 900                                                                       & 900                                                                        & 112        & Field corrector, slit mask               \\
\textbf{Wavelength range  {[}dz{]}}          & 150\AA{} {[}0.12{]}     & 75\AA{} {[}0.06{]}                                                            & 75\AA{} {[}0.06{]}                                                             & 0.5        & Emphasize high  $t_{atm}$ band          \\
\textbf{\# of CGM region/field}          & 1                   & 50                                                                        & 50                                                                         & 50         & IFU $\rightarrow$  Multi slit mask                     \\
\textbf{Selection}                           & Blind               & Preselected                                                               & Preselected                                                                &            &                                         \\
\hline

\textbf{Sensitivity (2hrs, $30''$)}            & 74,000 LU      & $\sim$ 80,000 LU  & $\sim$8,000 LU                                                                   & 9         & Direct detection ($5\sigma$)            \\

\hline
                 
\end{tabular}

\caption{Evolution of FIREBall scientific performances. }{Notes: $\alpha$: Improvement ratio between 2009 FB-1 flight and 2021 FB-2 flight. $\beta$: Net effi. = Effective QE $\times$ Instr. eff $\times$ Atm trans.}
\label{Improvements}

\end{table*}

\subsection{Comparison with existing observations}
For total flux (unresolved source), according to \textit{GALEX} observations and simulations (Figure 7 of \textit{Augustin et al. (2019)} \cite{Augustin2019}), the $\sim 41.7$ $
log(L_{Ly\alpha})$ limit we find for both \textit{scenarios} in Section \ref{sec:snr} would be constraining by an order of magnitude for the Ly$\alpha$ integrated flux at $z\sim0.7$. Stacking 10 objects per flight with similar galaxy properties/morphologies, using a very accurate continuum subtraction, should decrease this limit to $\sim 41.1$.\\

Concerning diffuse CGM emission, observations at higher redshift with \textit{MUSE} \cite{Wisotzki2015} revealed some Ly$\alpha$ haloes at $z\sim3$ reaching $4''$ ($\sim 30$ kpc) from the galaxy at their limiting surface brightness of $\sim 10^{-19}$ erg s$^{-1}$cm$^{2}$ arcsec$^{-2}$.\footnote{The discussion in \textit{Sullivan et al. (2019)}\cite{OSullivan2019} suggests that average $Ly\alpha$ surface brightness in the vicinity of QSOs might decrease with time (study from $z\sim3$ to $z\sim2$).}  With the assumed angular resolution of $5''$ for the 2021 flight, the first FIREBall-2 resolution element after the center is already at $\sim 35$ kpc at $z\sim0.7$. The same physical distance at $z\sim3$ ( $4.4''$) is just beyond \textit{Wisotzki et al. (2015)}\cite{Wisotzki2015} published radius.

For a spectrally unresolved object, the favorable $\sim \times30$ cosmological emission line surface brightness dimming at $z\sim3$ with respect to $0.7$ 
makes the FIREBall-2 single-object limit translate to $\sim 14.7 \times 10^{-19}$ erg s$^{-1}$cm$^{2}$ arcsec$^{-2}$ at $z\sim3$, a value five times brighter than MUSE's \cite{Wisotzki2015}.
Stacking 25 or more galaxies \cite{Augustin2019} is thus expected to provide a detection surface brightness limit at $z \sim 0.7$ comparable to that of MUSE at $z\sim3$ for individual galaxies.
FIREBall detection of extended emission would thus only be possible in presence of large spectral offsets due to strong winds.
Because of this, quasars, which have a brighter and more extended diffuse Ly$\alpha$ emission, are an even more interesting candidate for the 2021 flight \cite{Martin2014}.

\section{Conclusion}

The 2018 flight of FIREBall-2 performed the first space multi-object acquisition using a MOS, demonstrated the successful performance of all subsystems, and provided a test flight for several key technology innovations such as the EMCCD \cite{kyne2020deltadoped}, the sub-arcsecond pointing system\cite{Montel2019}, or the aspherical grating\cite{Lemaitre2014}. In this paper, we have described the calibration of the second generation of the FIREBall instrument and its performance. 

FIREBall-2 gained a factor $\sim 10$ in sensitivity compared to the first generation of the instrument (mostly detector and optical throughput) and now    has the capability to target $\sim50$ galaxies per mask. Unfortunately, the very high sky background on the detector due to the near full moon and the punctured balloon reflecting light to a non-baffled instrument part, combined with a shortened flight, wiped out this efficiency gain. A new baffling of the instrument is being implemented to mitigate stray light impact. Aside from this, only minor changes will be performed in order to launch FIREBall-2 again to minimize risks and ensure the full benefit of the factor 10 efficiency gain from generation one. 

The results of the 2018 flight combined with the ultimate sensitivity analysis validates the instrument design and its sub-systems. It demonstrates its capability to make valuable detections of the CGM and circum quasar medium and to provide critical spatio-spectral information about the medium surrounding $z\sim 0.7$ galaxies and QSOs. Such information is currently missing in our search for gas dynamics in the CGM at intermediate redshifts. FIREBall-2 is the only UV instrument currently proven and able to make these necessary discoveries at intermediate redshift (0.2$<z<$2). 

Once FIREBall-2 puts a constraint on CGM emission at $z\sim0.7$, it motivates more detailed follow-up with larger-class space missions, as they  would finally know the sensitivity, spatial, and spectral resolution constraints required to successfully map this emission. The implementation of this project is thus an investment for future orbital projects such as LUVOIR or MIDEX-class telescopes (HALO). 
This experiment, developed in the limited range of the stratospheric flights budget, allows both innovation with higher risk and low cost iterative improvements. 

\acknowledgments 
We acknowledge CNES and CNRS for supporting the French side of the FIREBall collaboration and the NASA grant NNX17AC56G obtained through the APRA program, which supports the FIREBall-2 mission in the US.
Vincent Picouet acknowledges CNES and NASA for the funding of his PhD at Aix-Marseille Universite and Columbia University. Ground support was provided by the Columbia Scientific Balloon Facility (CSBF) during the Fall 2017/2018 FIREBall-2 campaigns. 
The JPL detector team gratefully acknowledge the collaborative effort with Teledyne-e2v. This work is performed in part at Jet Propulsion Laboratory, California Institute of Technology under a contract with NASA.
\newpage

\bibliography{library}   
\bibliographystyle{spiejour}   

\end{multicols}

\listoffigures
\listoftables

\end{spacing}
\end{document}